\documentclass[floatfix,reprint,pre,aps]{revtex4-1}

\usepackage{amssymb} 
\usepackage{amsmath} %
\usepackage{natbib}
\usepackage{booktabs} 
\usepackage{caption}
\usepackage{subcaption}

\usepackage{graphicx} 

\graphicspath{ {figures/} } 

\begin{document}
\title{The effects of inhibitory and excitatory neurons on the dynamics and control of avalanching neural networks}
\author{Jacob Carroll}
\affiliation{Department of Physics (MC 0435) and Center for Soft Matter and Biological Physics, Virginia Tech, Robeson Hall, 850 West Campus Drive, Blacksburg, Virginia 24061, USA}
\email{jac21934@vt.edu}
\email{tauber@vt.edu}
\author{Ada Warren}
\affiliation{Department of Physics (MC 0435) and Center for Soft Matter and Biological Physics, Virginia Tech, Robeson Hall, 850 West Campus Drive, Blacksburg, Virginia 24061, USA}
\author{Uwe C. T{\"a}uber}
\affiliation{Department of Physics (MC 0435) and Center for Soft Matter and Biological Physics, Virginia Tech, Robeson Hall, 850 West Campus Drive, Blacksburg, Virginia 24061, USA}
\begin{abstract}
  The statistical analysis of the collective neural activity known as avalanches provides insight into the proper behavior of brains across many species. We consider a neural network model based on the work of Lombardi, Herrmann, De Arcangelis et al. that captures the relevant dynamics of neural avalanches, and we show how tuning the fraction of inhibitory neurons in this model alters the connectivity of the network over time, removes exponential cut-offs present in the distributions of avalanche strength and duration, and transitions the power spectral density of the network into an ``epileptic'' regime. We propose that the brain operates away from this power law regime of low inhibitory fraction to protect itself from the dominating avalanches present in these extended distributions. We present control strategies that curtail these power law distributions through either random or, more effectively, targeted disabling of excitatory neurons. 
\end{abstract}

\maketitle

\section{Introduction}
Neurons are the ubiquitous cells found in all intelligent animals, whose operations and dynamics are responsible for cognition. Individually the dynamics of neurons are well known  \cite{Neuron_Biology,Human_Physiology,Gaiarsa2002,Gerrow2010,Malenka2004}, but neurons are connected dynamically in ensembles of billions, and the collective behavior of these networks of neurons that gives rise to high-level cognitive functions such as memory and consciousness is only partially understood.

A common method for analyzing the collective behavior of biological neural networks is through the study of neural avalanches. An avalanche is a period of continuous neural activity in the network where signals are continually transmitted from neuron to neuron. Distributions of various quantities of these neural avalanches, such as the avalanche strength, duration, and power spectral density, are observed to follow power laws, and the exponents of these power laws appear to govern the govern the proper functioning of the network \cite{Beggs_Plenz_2003,Yan2016,Yu2014}. 

In this paper we discuss a model based on the work of Lombardi, Herrmann, De Arcangelis et al. \cite{deArcangelis_BalanceExcInh, deArcangelis_DragonKing, deArcangelis_TemporalOrg,deArcangelis_Power_Spectrum} of an avalanching neural network, that correctly reproduces the power law behavior of the avalanche strength distribution, duration distribution, and power spectral density of neuron activity. We show how these power laws are affected by modifying the fraction of inhibitory neurons, neurons that serve to suppress signals in the network, and observe intriguing extended power law behavior indicative of criticality in the avalanche strength and duration distributions, as well as exponents suggestive of epileptic behavior in the power spectral density at low inhibitory fractions. We also monitor how the outgoing connectivity distribution of the the network evolves under the effects of different inhibitory fractions.

Finally, we present two distinct strategies to control and remove the extended tails of the avalanche strength and duration  distributions in networks with low inhibitory fractions through the disabling of either random or highly connected neurons. Removing these extended tails serves to protect these networks from the extreme avalanches that occur in these extended distributions.

\subsection{Neurons}
\begin{figure}[htp]
  \centering
  \includegraphics[width=.5\textwidth]{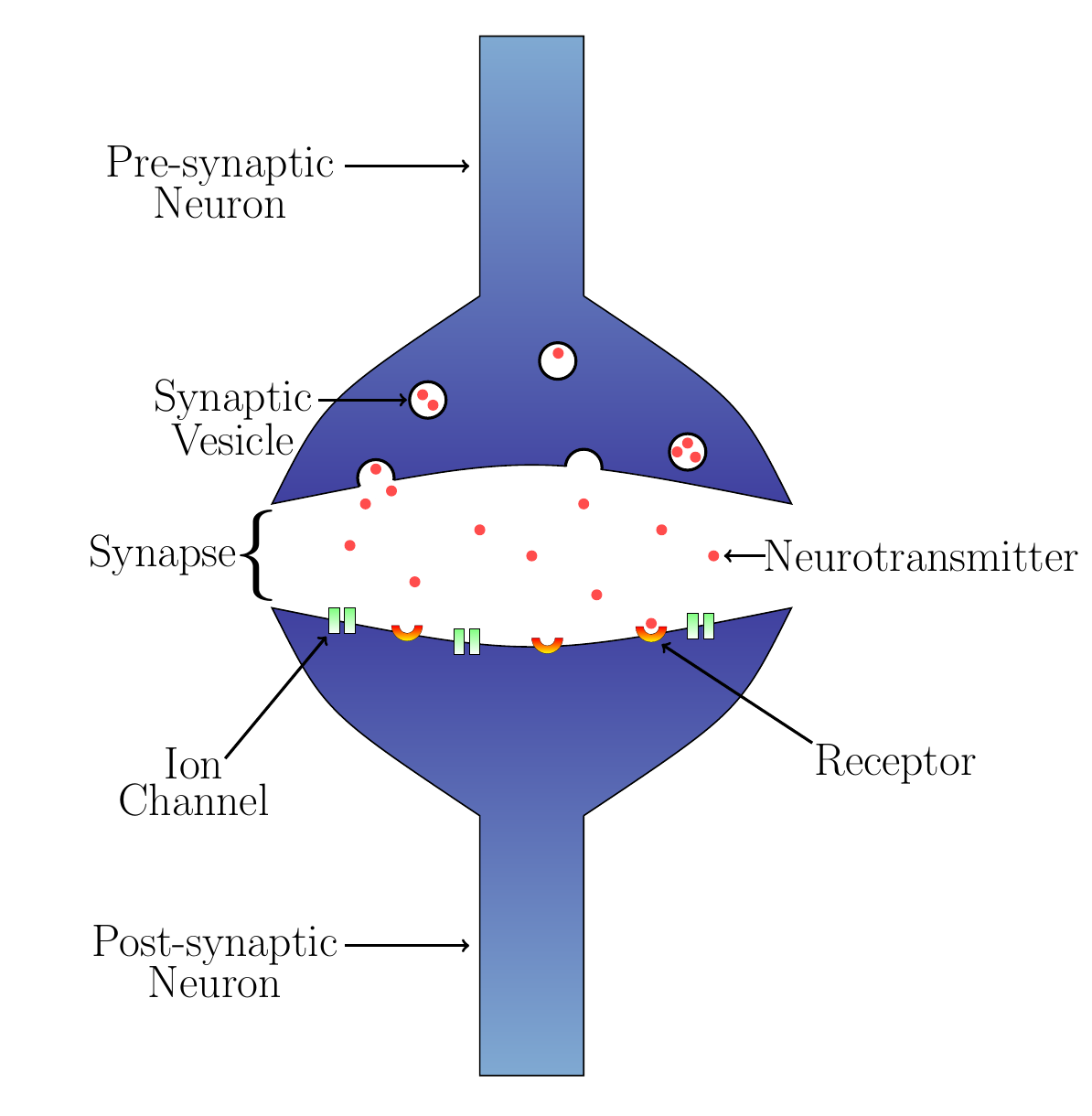}
  \vspace{-.5cm}
  \caption{\label{fig:synapse} A schematic of a synapse. The upper half of the image represents the pre-synaptic neuron, while the lower half represents the post-synaptic neuron. The gap between the pre- and post-synaptic neurons is the synapse. Neurotransmitters (small red circles) are collected in the synaptic vesicles (circular cavities in the pre-synaptic neuron). When the neuron fires, the synaptic vesicles are transported to the surface of the neuron at the synapse and expel their neurotransmitters. The neurotransmitters propagate through the synapse and bind to receptors on the surface of the post-synaptic neuron (red-orange half-circles). These receptors activate ion channels (paired green rectangles) on the post-synaptic neuron which cause the internal potential of the post-synaptic neuron to change as ions are transported into or out of the post-synaptic neuron's cell body \cite{Neuron_Biology, Human_Physiology}.}
\end{figure}

Biological neural networks are composed of individual neurons connected to each other by synapses. Synapses are small gaps between neighboring neurons where neurons can release and receive neurotransmitters: chemicals that cause the receiving neuron to open or close ion gates and pumps in order to increase or decrease its membrane potential, depending on the neurotransmitter received. Some neurons solely release inhibiting neurotransmitters, and will be referred to as ``inhibitory neurons". These neurons serve to suppress signals in brain, and make up 20-30\% of the neurons in the human cortex \cite{deArcangelis_Power_Spectrum,Sahara2012}. As a matter of definition, the neuron releasing the neurotransmitters will be referred to as the ``pre-synaptic" neuron, and the neuron receiving the neurotransmitters will be referred to as the ``post-synaptic" neuron \cite{Neuron_Biology}. A schematic of a synapse is shown in Fig. \ref{fig:synapse}.

If the membrane potential of the post-synaptic neuron is increased beyond a threshold value, $-55 mV$, then the post-synaptic neuron generates a spiking potential that travels down the length of the neuron's cell body. Synapses that this signal reaches will release neurotransmitters of their own, to be picked up by other neurons. This signal is referred to as an ``action potential", and this process will summarily be called ``firing." It is important to note that the connections between neurons are not symmetric. It is not necessarily true that if neuron A can transmit to neuron B, then neuron B can transmit to neuron A \cite{Neuron_Biology}.

Pre-synaptic neurons that have just fired lock down their ion channels until they can recoup the resources used in generating an action potential and releasing neurotransmitters, for a period of 3-4ms \cite{Human_Physiology}. During this period, the neuron cannot respond to any received neurotransmitter, and is for our purposes dormant. This period is referred to as the ``refractory period."

The strength of signals transmitted across synapses is determined by the combination of several different factors from both the pre- and post-synaptic neurons, such as the quantity of neurotransmitters produced by the pre-synaptic neuron, and the number of neurotransmitter receptors available on the post-synaptic neuron's surface. The strength of these signals can change over time \cite{Gaiarsa2002,Gerrow2010,Malenka2004}.

Synapses that successfully perpetuate signals, i.e., when the pre-synaptic neuron causes the post-synaptic neuron to fire, increase their efficiency through the increased production of neurotransmitters in the pre-synaptic neuron \cite{Gaiarsa2002} or an increase in the number of receptors on the post-synaptic neurons \cite{Gerrow2010}. Synapses that do not perpetuate signals have their neurotransmitter production or receptor number reduced \cite{Malenka2004}. This activity-dependent change in synapse strength is known as Hebbian learning \cite{Cooper_Hebb}, and is a feature the model was devised in Refs. \cite{deArcangelis_DragonKing,deArcangelis_Power_Spectrum,deArcangelis_BalanceExcInh,deArcangelis_TemporalOrg} to capture.
\subsection{Avalanches}
In the context of neural networks, avalanches are defined as a period of continuous neural activity. The name is chosen in analogy to Abelian sandpile models, where a column of sand can topple sending sand down onto lower columns causing them to topple and so on, creating a literal avalanche of sand \cite{Sandpile}. The same process can happen among neurons where a single neuron fires, causing other neurons to fire, et cetera, until the network has temporarily exhausted its resources. The period of continuous neuron firing is an avalanche.     

The dynamics of neural avalanches were studied by Beggs and Plenz in rat cortex cultures, and various properties of avalanches in these cultures were found to follow power law distributions \cite{Beggs_Plenz_2003}. The observables investigated included the avalanche strength: the sum of all signals sent by firing neurons; and the avalanche duration: the length of time that the avalanche persists. These results for avalanche strength have since in 2014 been replicated in macaque monkeys  \cite{Yu2014}. 

Additionally, power law behavior has been recorded in the power spectral density of neural activity in humans using electroencephalography and electrocorticography \cite{Yan2016}, and avalanche models have been shown to replicate this behavior, even matching exponents seen experimentally \cite{deArcangelis_BalanceExcInh}.  

\section{Neural network model} 
\label{section:model}
Our numerical model is based on the work of Lombardi, Herrmann, De Arcangelis et al. \cite{deArcangelis_BalanceExcInh, deArcangelis_DragonKing, deArcangelis_TemporalOrg,deArcangelis_Power_Spectrum}, and in addition to modeling key features of biological neural networks; namely firing at a threshold potential, refractory periods, and Hebbian learning \cite{Cooper_Hebb}, the model accurately reproduces several experimentally determined distributions related to avalanches of neural activity in biological neural networks \cite{Beggs_Plenz_2003, Yan2016}.

An extension to this model developed by Lombardi, Herrmann, De Arcangelis et al. which in addition recreates avalanche waiting time distributions is described in the Appendix.

The following simplified model variant however does not attempt to recreate these waiting time distributions, and this extension is excluded to reduce the complexity of the system. Table \ref{tab:network_params} lists the various parameters used in our model.  

\begingroup
\squeezetable
\begin{table}[htp]
  \caption{\label{tab:network_params}The network parameters used in the avalanching neural network model.}
  \begin{ruledtabular}
    \begin{tabular}{@{}llll}
      Parameter &  Description & Value \\
      \hline
      $N$ & Number of neurons in the network & $64000$ & \\
      $t$ & Time step & 10$ms$ \\
      $p_{inh}$ & Fraction of inhibitory neurons in the network & --- \\
      $J_{ij}(t)$ & Weight of connection from $i^{\text{th}}$ to $j^{\text{th}}$ neuron & --- \\
      $g_{ij}(t)$ & Weight scaled by degrees of connectivity & --- \\
      $J_{min}$ & Minimum weight strength & 0.001 \\
      $J_{max}$ & Maximum weight strength & 2 \\
      $n_{i}(t)$ & Potential of $i^\text{th}$ neuron & ---\\
      $s_{i}(t)$ & Action potential of $i^\text{th}$ neuron & --- \\
      $n_{\text{max}}$ & Threshold potential & -55$mV$\\
      $k_{{in}_i}(t)$ & Number of incoming connections to $i^{\text{th}}$ neuron & ---  \\
      $k_{{out}_i}(t)$ & Number of outgoing connections from $i^{\text{th}}$ neuron & ---  \\
    \end{tabular}
  \end{ruledtabular}
\end{table}
\endgroup

\subsection{Neuron dynamics}
\label{Neuron_dynamics}
The model consists of a number of neurons $N$, each neuron $i$ defined by its potential $n_i$. At the beginning of the simulation, each neuron is randomly designated as inhibitory, with probability $p_{inh}$, or excitatory with probability $1-p_{inh}$. This will determine whether signals from this neuron increase (excitatory) or decrease (inhibitory) the potentials of other neurons.

The neuron is then randomly assigned an out-degree $k_{{out}_i}$ from a truncated power law distribution, formed such that $P(k_{out})\sim k_{out}^{-2}$ for $k_{out}\in [2, 100]$. This range was chosen to mimic the distribution of connectivity experimentally observed in human cortices, which demonstrates power law behavior across two decades of connectivity, following an exponent of -2 \cite{Eguiluz2005,Lee2010}. Additionally, the truncated nature of the power law also allows this distribution to be normalized. The $k_{{out}_i}$ neurons are then chosen from a uniform distribution, and connections between them and the $i^{\text{th}}$ neuron are established by assigning each synapse an initial weight $J_{ij}$ uniformly distributed on the interval $(0, 1)$. Connections from a neuron to itself are not allowed in the model. Once the initial network topology is established, the in-degree $k_{{in}_i}$ is tabulated for each neuron.

The network is initialized such that the potential of each neuron is set to 90\% of the threshold potential $n_{max} \sim -55mV$ to facilitate and accelerate the initial building-up of network activity.

During each time step $t$, any neuron whose potential has increased past the system's firing threshold, $n_i(t) \geq n_{max}$, fires, sending an action potential $s_{i}(t)$ to each of the $k_{{out}_i}(t)$ connected neurons. If the potential of the  $i^{\text{th}}$ neuron is not above the threshold potential, the action potential is zero: 
\begin{equation}
  \label{eq:action_potential}
  s_{i}(t) =
  \begin{cases} 
    0, &  n_{i}(t) < n_{max}\\
    n_{i}(t), & n_{i}(t) \geq n_{max}\; .
  \end{cases} 
\end{equation}

After a neuron has fired, its potential is set to zero for one time step, during which it cannot receive signals from other neurons. This mimics the refractory period of real neurons \cite{Neuron_Biology}. The signal received by the post-synaptic neurons not in a refractory period is proportional to the the action potential of the pre-synaptic neuron. This behavior is summarized in Eq. (\ref{eq:fire}),
\begin{equation}
  \label{eq:fire}
  n_{j}(t+1) =
  \begin{cases}
    0, & s_{j}(t) > 0\\
    n_j(t) \pm g_{ij}(t) s_i(t),  & s_{j}(t) = 0 \; ,  \end{cases}
\end{equation}
where the upper and lower signs are for $i$ excitatory and inhibitory, respectively, and $g_{ij}(t)$ controls how much the $j^{\text{th}}$ neuron is affected by signals from the $i^{\text{th}}$ neuron:
\begin{equation}
  g_{ij}(t) = \frac{ k_{{out}_i}(t)}{k_{{in}_j}(t)} \frac{J_{ij}(t)}{\sum_k J_{ik}(t)}\; .
\end{equation}

While the strength of the synaptic signal is proportional to the pre-synaptic neuron's action potential potential, this signal is scaled by a factor $J_{ij}(t)/\sum_{k}J_{ik}(t)$ that determines the relative strength of the connection between the $i^\text{th}$ and $j^\text{th}$ neurons compared to all connections from the $i^\text{th}$ neuron. This is then in turn scaled by the factor $k_{out_i}(t)/k_{in_j}(t)$. 

While $J_{ij}(t)/\sum_{k}J_{ik}(t)$ determines the strength of the $i\rightarrow j$ connection relative to all of the  $i^\text{th}$ neuron's connections, the factor $k_{out_i}(t)/k_{in_j}(t)$ rescales each of these connections relative to their importance to the network. This is necessary in order to properly compare the strength of signals from different neurons.

A neuron with many outgoing connections will be more important to the network than a neuron with few, and so its outgoing signals will be scaled by the neuron's number of outgoing connections, $k_{out_i}(t)$. A neuron that receives signals from many different neurons will be less excited by any one connection, so any incoming signals to it will be scaled by its number of incoming connections, $k_{in_j}(t)$. Indeed, this factor $k_{out_i}(t)/k_{in_j}(t)$ is responsible for the power law distribution of the avalanche strengths, though it has no bearing on the waiting time or duration distributions.
The change in potential in Eq. (\ref{eq:fire}) is referred to as the depolarization of $j$ due to $i$ for this time step. 

A series of successive time steps during each of which at least one neuron fires constitutes an avalanche.

\begin{figure}[htp]
  \centering	
  \includegraphics[width=.5\textwidth]{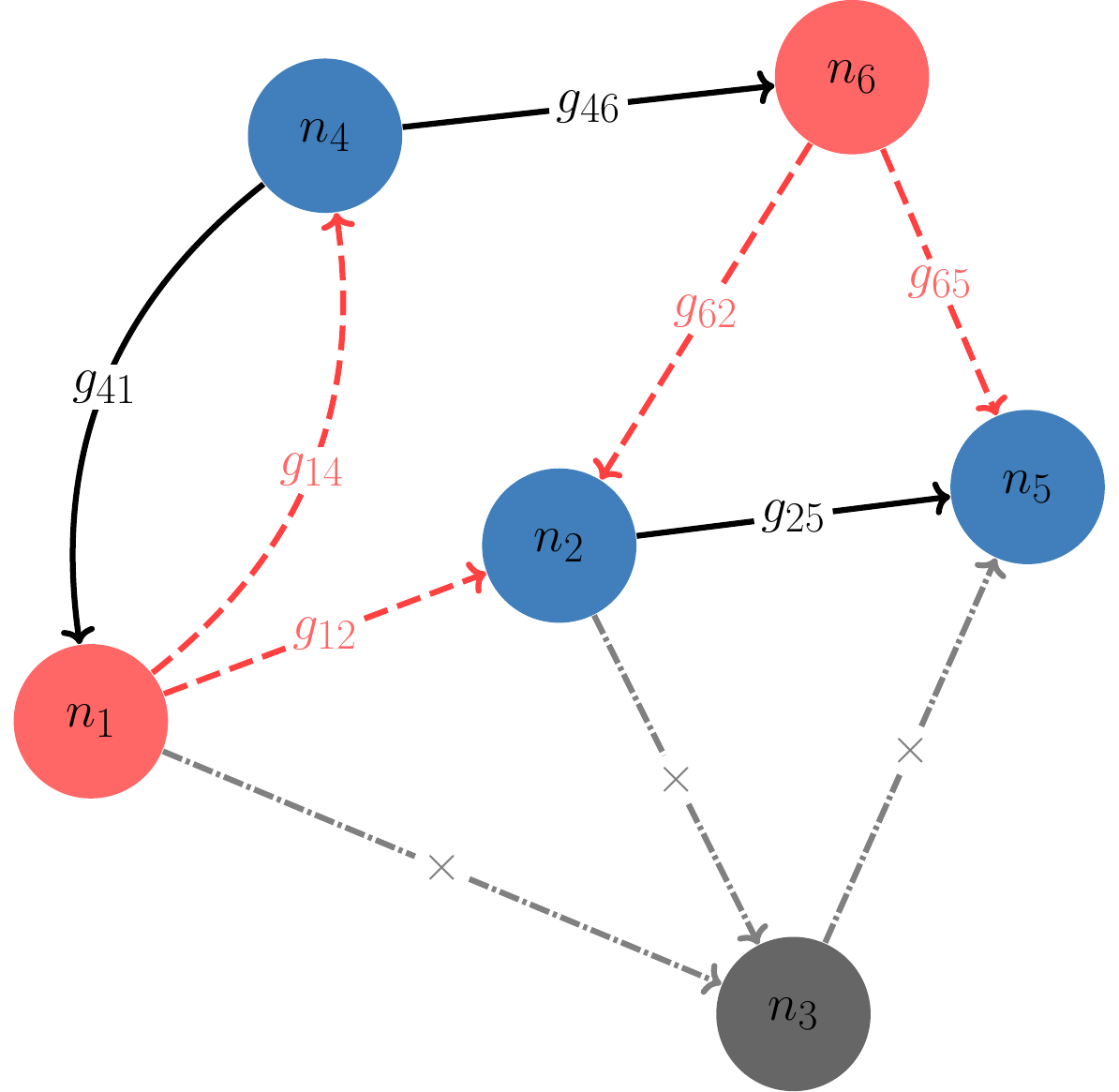}
  \caption{\label{fig:example_network} An example network of six neurons showcasing the various possible interactions between neurons. The colored circles represent six different neurons, labeled $n_{i}$, while the arrows connecting them represent the synaptic connections between neurons with the corresponding elements of the weight matrix $g$ centered in each line. The (blue) neurons $n_2$, $n_4$, and $n_5$ have zero action potential and send no signals through their connections (the black lines) but receive signals from firing neurons. The (red) neurons $n_1$ and $n_6$ are firing, sending their non-zero action potentials through their outgoing connections (the red dashed lines). The (grey) neuron $n_3$ is in a refractory period, and can neither send or receive signals through its synaptic connections (the grey dash-dotted lines).}
\end{figure}

Figure \ref{fig:example_network} shows a schematic of a simple, six neuron network with examples of different possible connections and neuron behavior.

\subsection{Hebbian learning and pruning}
After each avalanche, the strength of connections between neurons is adjusted according to Hebbian-like rules \cite{Cooper_Hebb}, and then pruned (set to zero) if the strength of connection drops below a threshold $J_{min}$.

Due to the variable length of each avalanche, it is convenient to index avalanches on a separate variable $\tau$ where each avalanche has beginning and ending times $t_{i}(\tau)$ and $t_{f}(\tau)$. At the end of each $\tau^{\text{th}}$ avalanche, we implement Hebbian-like plasticity rules. The strength of each synapse $J_{ij}$ is increased proportional to the sum of all signals sent through the synapse during the avalanche, and decreased by the average increase in synaptic strength. We cap each $J_{ij}$ to a maximum value of $J_{max}$ in order to ensure stability in the network. The change in synaptic strength is summarized in Eq. \ref{eq:hebbian}:
\begin{equation} 
  \label{eq:hebbian}
  J_{ij}(\tau) =	J_{ij}(\tau - 1) + \frac{\delta n_{ij}(\tau)}{n_{max}} - \Delta J(\tau)\; ,
\end{equation}
where $\delta n_{ij}(\tau)$ is the sum of magnitudes of all signals sent from neuron $i$ to neuron $j$ during the $\tau^{th}$ avalanche,

\begin{equation}
  \label{eq:delta_n}
  \delta n_{ij}(\tau) = \sum_{t=t_{i}(\tau)}^{t_{f}(\tau)} \frac{ k_{{out}_i}(t)}{k_{{in}_j}(t)} \frac{J_{ij}(t)}{\sum_k J_{ik}(t)}s_i(t)\; ,
\end{equation}
and $\Delta J(\tau)$ is the average increase in connection strength after the $\tau^{\text{th}}$ avalanche, 
\begin{equation} \label{eq:reduce}
  \Delta J(\tau) = \frac{1}{N_{C}(\tau)}\sum_{i=1}^{N}\sum_{j=1}^{N}\frac{\delta n_{ij}(\tau)}{n_{max}}\; .
\end{equation}
Here, $N_{C}(\tau)$ is the number of non-zero connections in the network,
\begin{equation*}
  N_{C}(\tau) = \sum_{i,j}\Theta(J_{ij}(\tau))\; ,
\end{equation*}
where $\Theta$ represents Heaviside's step function. The competition between the second and third terms of Eq. (\ref{eq:hebbian}) causes used connections to increase in strength, while unused connections weaken. If any connection is lowered below a threshold $J_{min}$, the connection is permanently removed as that element $J_{ij}$ is set to zero. This removal of weak connections is called pruning.

To prevent over-pruning, we impose an upper bound of $J_{max}$ on the strength of any given connection. This strengthening procedure can saturate this bound, but not exceed it. This rule forces the network to prioritize those connections which are most often used, as in biological neural networks \cite{Cooper_Hebb}. 

Between avalanches, the system is stimulated via small ($\sim1\%$ of the threshold potential) constant noise applied to randomly chosen neuron potentials. This noise tends to drive the system towards another avalanche.

\section{Distributions of avalanche parameters}
\label{Distributions}
Several different parameters of neural avalanches can be studied through statistical analysis, and have been shown to be important to the proper operation of biological neural networks \cite{Beggs_Plenz_2003,Yu2014,Yan2016,Myers2017}. These observables are the avalanche strength, the avalanche duration, and the power spectral density of neuron activity. In this section we describe each of these quantities in turn, and our methods for modeling and recording them.
\subsection{Avalanche strength}
\label{Avalanche Strength}
The strength of the $\tau^\text{th}$ avalanche is defined to be the sum of absolute values of all signals sent between neurons during the avalanche. 

Biologically, this is the sum of all neuron action potentials. This has been recorded experimentally through the careful placement of electrodes on both {\sl in-vivo} and {\sl in-vitro} neural networks \cite{Beggs_Plenz_2003,Yu2014}. The strength of many different avalanches can be collected to form a probability distribution describing the likelihood of a given avalanche having a certain strength $P_{S}$, where $S$ is the strength of a given avalanche.
This distribution $P_{S}$ has been found to follow a power law of $P_{S}(S)\sim S^{-1.5}$ \cite{Beggs_Plenz_2003,Yu2014}.

We calculate the total avalanche strength by summing the strength of each signal sent between neurons during an avalanche. If each avalanche has beginning and ending times $t_{i}(\tau)$ and $t_{f}(\tau)$ respectively, we define the strength of the avalanche as:

\begin{equation}
  \label{eq:ave_strength}
  S(\tau) = \sum_{t=t_{i}(\tau)}^{t_{f}(\tau)}\sum_{i=1}^{N}\sum_{j=1}^{N} \frac{ k_{{out}_i}(t)}{k_{{in}_j}(t)} \frac{J_{ij}(t)}{\sum_k J_{ik}(t)} s_i(t)\; .
\end{equation}
\subsection{Avalanche duration distribution}
The duration of an avalanche is the length of time that the avalanche persists. This has been recorded experimentally in the same manner as the avalanche strength, and has been collected into a distribution $P_{D}$ describing the likelihood of a given avalanche duration.  
The duration of the $\tau^{\text{th}}$ avalanche, $D(\tau)$, is taken to be the number of time steps that the avalanche persists for. This can be written as

\begin{equation}
  \label{eq:duration}
  D(\tau) = t_{f}(\tau) - t_{i}(\tau)\; ,
\end{equation}
where $t_{i}(\tau)$ is the initial time step of the $\tau^{\text{th}}$ avalanche, and $t_{f}(\tau)$ is the final time step of the $\tau^{\text{th}}$ avalanche.

\subsection{Avalanche power spectral density}
The power spectral density (PSD) of a signal describes the distribution of power in the signal as a function of frequency. This is a common form of analysis done on the measurements of {\sl in-vivo} neural activity via techniques such as electroencephalography and electrocorticography. Electroencephalography and electrocorticography both measure the action potentials of neurons firing in living brains through the placement of electrodes either outside (electroencephalography) or inside (electrocorticography) the skull. The time series measurement of electrical activity can be decomposed into their power spectral density and has been suggested as a means to diagnose epilepsy \cite{Yan2016,Myers2017}. Healthy non-epileptic brains have a PSD exponent in the range of $(-1.5,-0.8)$ \cite{deArcangelis_Power_Spectrum}, while brains undergoing epileptic events have been recorded with PSD exponents in the range of $(-2.2,-1.8)$ \cite{Yan2016, deArcangelis_Power_Spectrum}.

In our simulations the sum of all depolarizations is calculated after each time step in an avalanche. This sum is then appended to a time series as $x_{n}$, the $n^{th}$ sum of depolarizations. We perform this summation for each time step in every avalanche until the series $\{x_{n}\}$ contains the sum of depolarizations for every time step of every avalanche.  

The power spectral density of time series data can be determined by computing this series' discrete Fourier transform. The average of the square of the contribution of each frequency in the discrete Fourier transform gives the power spectral density of that frequency.

\begin{equation}
  \label{eq:PSD}
  PSD(f) = \frac{1}{N}\Bigg|\sum_{n=1}^{N}x_{n}e^{-i f n}\Bigg|^{2},
\end{equation}
where PSD is the power spectral density as a function of the frequency $f$, $x_{n}$ is the sum of depolarizations in the $n^{th}$ time step, and $N$ is the total number of time steps that these depolarizations were recorded for. We compute the PSD across the entire frequency range using Welch's method and record it in a histogram using logarithmic binning to smooth out fluctuations that occur at the lower frequency ranges due to their limited occurrence in our data. This limited occurrence is due to the power law nature of the model's PSD (see Fig. \ref{fig:More_PSD}).  

\section{Results}
For each of the following results we simulated 100 different networks of 64,000 neurons, each randomly initialized according to the methods described in Sec. \ref{Neuron_dynamics}.
Each network was allowed to operate for $10,000-100,000$ separate avalanches, after which the various distributions described in Sec. \ref{Distributions} were calculated and averaged across the 100 separate networks. The averaged distributions are recorded below. 

\subsection{Avalanche duration distribution}
\begin{figure}[htp]
  \centering
  \includegraphics[width=.5\textwidth]{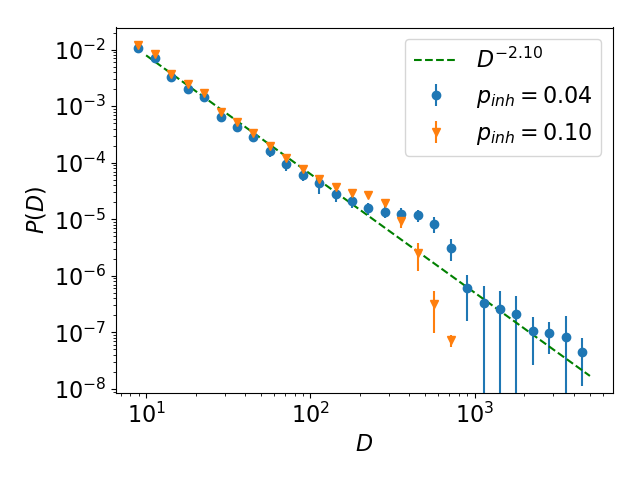}
  \caption{\label{fig:duration} Avalanche duration distributions for two 64,000 neuron networks with different inhibitory fractions $p_{inh}$. The x axis represents the duration $D$ of an avalanche in time-steps of $10ms$. The y axis represents the probability of this avalanche duration. The (blue) dots represent the distribution of avalanche durations with $p_{inh}=0.04$. The (orange) triangles represent the distribution of avalanche durations with $p_{inh}=0.1$. The (green) dashed line indicates a power law with exponent $-2.1$. As with the avalanche strength distributions (see Fig. \ref{fig:strength}) the exponential cut-off seen at $D=3\cdot10^{2}$ time-steps when $p_{inh}=0.1$ disappears and the power law behavior persists for several decades at the same exponent, before the statistics of the events become to poor. This extended tail for the lower inhibitory fraction ultimately dominates the dynamics of the system, as the avalanches in this regime last for orders of magnitude more time steps than the original regime. This tail continues for several more decades, but these data points have been trimmed from the plot due to poor statistics. The hump seen in both the high and low inhibitory fraction data around $D\sim 5\cdot 10^2$ is a finite-size effect related to the duration necessary for all neurons in the network to fire, on average, once. }
\end{figure}

Using the network described in Sec. \ref{Neuron_dynamics} and Eq. (\ref{eq:duration}) we measured the distribution histogram of avalanche durations. Figure \ref{fig:duration} shows the avalanche duration distributions for two networks of 64,000 neurons. The (blue) dots represent the distribution of a network with inhibitory fraction $p_{inh}=0.10$, while the (orange) triangles represent the distribution of a network with inhibitory fraction $p_{inh}=0.04$. The (green) dashed line indicates a power law with exponent $-2.1$.

For short avalanche durations both distributions agree well with the experimental results for the avalanche duration distribution which follows a power law with exponent of $-2.0$ followed by an exponential cut-off. The distribution with higher inhibitory fraction, $p_{inh}=0.10$, matches the experimental results very well, while the distribution from the network with low inhibitory fraction $p_{inh}=0.04$ does not show the experimentally found exponential cut-off, and instead displays continued power law behavior at the same exponent for several more decades before statistics of the measured events becomes too poor. This increase in available avalanche durations is due to the inability of the network to suppress signals because there are few inhibitory neurons in the network. This generates very long lasting avalanches that in turn give rise to the extended power law regime seen in the avalanche strength distribution shown in Fig. \ref{fig:strength}. These long-lasting avalanches ultimately dominate the dynamics of the network, because while they are rare, these avalanches can be up to $10^4$ time steps longer than avalanches seen in a network with a higher inhibitory fraction. 

The inhibitory fraction at which we see these long lasting avalanches occur is much lower than the value observed in human cortices, which is around $0.2-0.3$ \cite{Sahara2012,deArcangelis_Power_Spectrum}. Milton et al. suggest that the exponential cut-offs seen in the avalanche distributions exist to protect the brain from runaway avalanches; our results corroborate this idea that the brain might operate away from this regime in order to not be dominated by the incredibly long lasting avalanches present at low inhibitory fractions, so the brain actually ultimately avoids truly critical behavior.  

The hump displayed in both sets of data is a finite-size effect related to the duration necessary for a sufficiently strong avalanche to propagate through the entire network. Because the network is stimulated by small constant noise between avalanches, every neuron in the network will, on average, be very close to firing when an avalanche begins. This allows avalanches to initially propagate more easily through the network, until every neuron has fired at least once. After this, this initial ``supply" of neuron potential has been exhausted, and avalanches must be self sustaining to continue past this point. Many avalanches are not strong enough to continue propagating without many neurons in the network having highly elevated potentials, and so the avalanches end around this value of duration, creating a hump in the distribution. This finite-size effect is also apparent in the avalanche strength distribution, Fig. \ref{fig:strength}.

\subsection{Avalanche strength distribution}
\begin{figure}[htp]
  \centering
  \includegraphics[width=.5\textwidth]{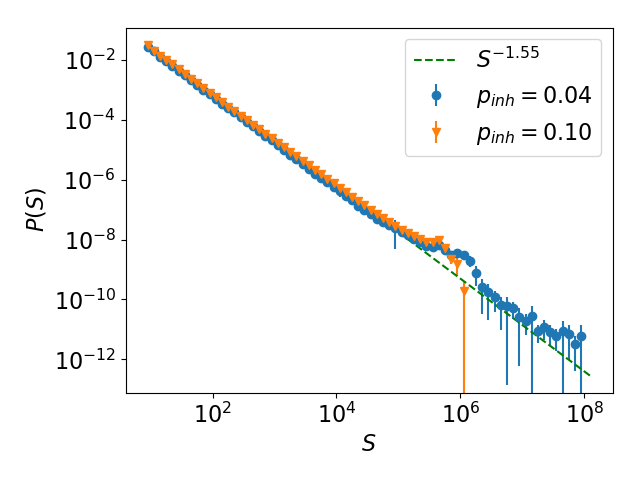}
  \caption{\label{fig:strength} Avalanche strength distributions for two 64,000 neuron networks with differing inhibitory fractions $p_{inh}$. All other network parameters are as described in Sec.  \ref{Neuron_dynamics}. The x axis represents the avalanche strength as defined in Sec. \ref{Avalanche Strength}. The y axis represents the probability of an avalanche occurring with that strength. The (blue) dots are the distribution of avalanche strength for a network with $p_{inh}=0.04$ inhibitory fraction. The (orange) triangles are the distribution of avalanche strength at $p_{inh}=0.1$. The (green) dashed line indicates a power law with exponent $-1.55$. Note that for $p_{inh}=0.04$ the exponential cut-off seen at $S=5\cdot 10^{5}$ in the $p_{inh}=0.1$ data disappears, and the power law behavior extends to much larger avalanche strengths. Avalanches were recorded with strengths up to $S\sim 10^{10}$, but these were excluded from the figure due to poor statistics. The bump seen at $S\sim 10^6$ in both distributions is a finite-size effect that is proportional to the simulation value of the neuron firing threshold multiplied by the number of neurons in the network. It forms because most neurons in the network are very close to firing when an avalanche begins, and this average increase in neuron potential helps sustain avalanches early on. Once every neuron in the network has fired, the avalanche must become self sustaining to continue. Many avalanches cannot continue propagating without the initial supply of potential, and die out when they have fired approximately all of the neurons in the network, causing this bump.}
\end{figure}

Using the network described in Sec. \ref{Neuron_dynamics} and Eq. (\ref{eq:ave_strength}) we measured the distribution histogram of avalanche strengths. Figure \ref{fig:strength} shows the avalanche strength distributions for two different networks, each made of 64,000 neurons. The two networks differ only in their inhibitory fraction $p_{inh}$. The (blue) dots represent the avalanche strength distribution of a network with $p_{inh}=0.04$ and the (orange) triangles represent the avalanche strength distribution of a network with $p_{inh}=0.10$. The (green) dashed line indicates a power law with exponent $-1.55$. In both cases the early avalanche strength distributions ($S\in [10^1, 10^5]$) follow a power law of $P_S\sim S^{-1.55}$, which agrees well with the distributions found in rat and macaque monkey cortices  \cite{Beggs_Plenz_2003,Yu2014}.

For larger inhibitory fractions, the power law behavior reaches an exponential cut-off at high avalanche strengths as the network is unable to sustain the activity necessary for massive avalanches.  Again, as seen in the avalanche duration distributions, as the inhibitory fraction decreases the network becomes better able to sustain increasingly larger avalanches, and at a inhibitory fraction of $p_{inh}=0.04$ we see the exponential cut-off disappear as the power law behavior is extended for several more decades. This extension of the power law behavior due to these massive avalanches suggests the system is approaching a critical regime as the inhibitory fraction is lowered.

Additionally, both the avalanche strength and duration distributions were found to be very stable with respect to the noise strength, minimum and maximum weight strengths, as well as the threshold potential, with changes in these parameters resulting in little to no modifications in their dynamics.

\subsection{Power spectral density}
\begin{figure}[htp]
  \centering
  \includegraphics[width=.5\textwidth]{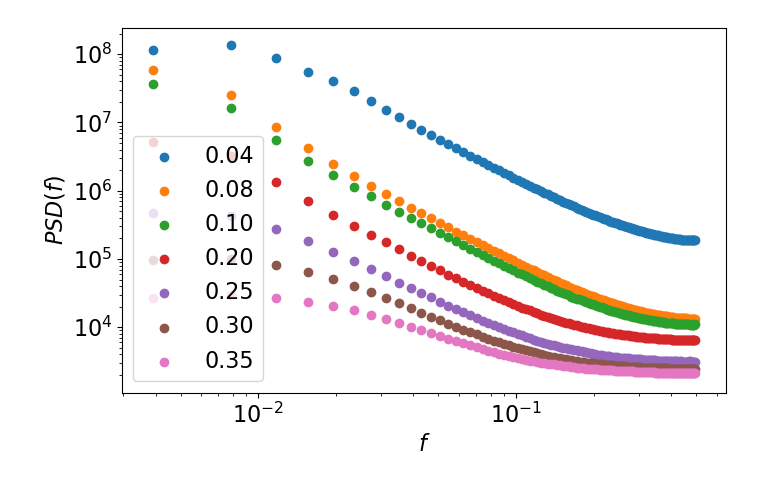}
  \caption{\label{fig:More_PSD} Seven different power spectral densities for 64,000 neuron networks with differing inhibitory fraction $p_{inh}$. As the inhibitory fraction is increased in the network, the average contribution across the entire frequency range decreases, and the power law behavior shifts from following an exponent of $-2.0$, which falls into the range of epileptic behavior for humans \cite{Yan2016,deArcangelis_Power_Spectrum} to $-1.6$ which is very close to the range of normal behavior for humans. These two power laws are shown in more detail in Fig. \ref{fig:PSD}. The PSD's displayed show two regimes of behavior: the aforementioned power law regime at lower frequencies, and a flattened semi-constant regime at high frequencies. The power law regime is due to long-range temporal correlations present between neuron firings. It is an assumption of this model that neurons firings are uncorrelated across different avalanches, so this power law regime becomes more pronounced  as the inhibitory fraction decreases, due to the increasing accessibility of long duration avalanches. For higher inhibitory fractions, there is an effective cut-off on the accessible durations which reduces the length at which these neuron firings can be temporally correlated, which causes an effective increase in the contribution due to higher frequencies. }
\end{figure}

\begin{figure}[htp]
  \centering
  \includegraphics[width=.5\textwidth]{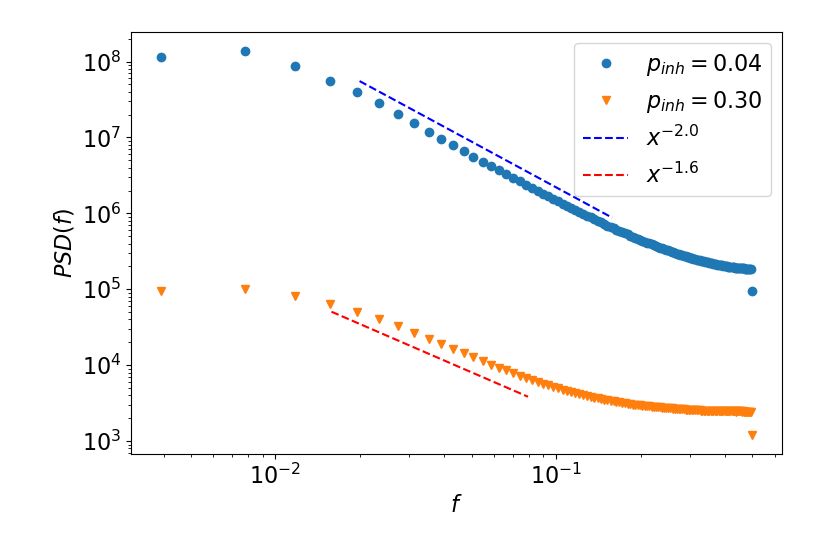}
  \caption{\label{fig:PSD} Two power spectral densities of 64,000 neuron networks with different inhibitory fractions $p_{inh}$. The (blue) dots represent the power spectral density of the network with $p_{inh}= 0.04$. The (orange) triangles represent the power spectral density of the network with $p_{inh}=0.30$. The (blue) dashed line indicates a power law with exponent $-2.0$, an exponent in  the regime of epileptic behavior seen in humans \cite{Yan2016, deArcangelis_Power_Spectrum}. The (red) dashed line indicates a power law with exponent $-1.6$, which is very close to the regime of normal operating brain behavior in humans \cite{Yan2016,deArcangelis_Power_Spectrum}.}
\end{figure}

Using Eq. (\ref{eq:PSD}) we constructed the power spectral density of our network of 64,000 neurons for a variety of inhibitory fractions. Figure \ref{fig:More_PSD} shows seven different network power spectral densities for inhibitory fractions ranging from $0.04$ to $0.35$. As the inhibitory fraction is increased, the average contribution of the entire frequency range decreases as the inhibitory neurons in the network suppress signals across all frequencies. Additionally the exponent of the power law the PSDs follow varies. This change is outlined in Fig. \ref{fig:PSD}. The PSD's shown in both Fig. \ref{fig:More_PSD} and Fig. \ref{fig:PSD} show two distinct dynamical regimes: a power law regime at low frequencies, and a semi-constant regime at high frequencies. The PSD's power law regime is due to long-range temporal correlations present between neuron firings. It is an assumption of this model that neurons firings are uncorrelated across different avalanches, so this power law regime becomes more pronounced as the inhibitory fraction decreases, due to the increasing accessibility of long duration avalanches. For higher inhibitory fractions, there is an effective cut-off on the accessible durations which reduces the length at which these neuron firings can be temporally correlated, which causes an effective increase in the contributions from higher frequencies and produces this flattening of the distribution at higher frequencies. 

Figure \ref{fig:PSD} shows the power spectral distribution for two networks with very high and very low inhibitory fraction, respectively. The (orange) triangles represent the PSD of the network with an inhibitory fraction $p_{inh}=0.30$, while the (blue) dots represent the PSD of the network with an inhibitory fraction $p_{inh}=0.04$. The (blue) dashed line represents a power law with exponent $-2.0$, while the (red) dashed line indicates a power law with exponent $-1.6$. 

The PSD of the low inhibitory network follows a power law with an exponent that is within the regime of epileptic behavior measured in human PSDs \cite{Yan2016,deArcangelis_Power_Spectrum}, while the network with a higher inhibitory fraction has an exponent very close to the regime of normal operating brain behavior for human PSDs \cite{Yan2016,deArcangelis_Power_Spectrum}. The inhibitory fraction of the higher network is also much closer to the inhibitory fraction found in human cortices \cite{deArcangelis_Power_Spectrum,Sahara2012}.

We observe these epileptic exponents in the PSD in the same regime of inhibitory fraction where we see the extended power laws in the distributions for avalanche strength and duration. 

\subsection{Neuron connectivity distribution}
\begin{figure}
  \includegraphics[width=.5\textwidth]{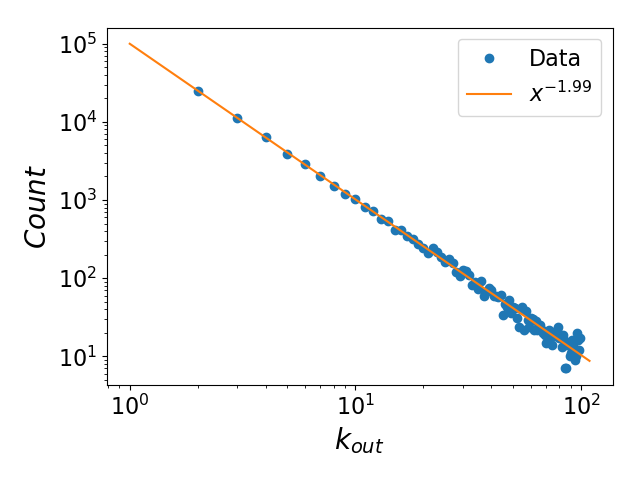}
  \caption{\label{fig:init} An example of the initial $k_{out}$ degree distribution for our simulations. The  horizontal axis represents the number of outgoing connections $k_{out}$ a neuron has, while the vertical axis represents the number of neurons in the network with a particular $k_{out}$. The (blue) circles represent measured degrees of connectivity for all the neurons in a network upon initialization, while the (orange) line displays the best fit of a power law to the data.  As described in Sec. \ref{section:model}, the outgoing degrees of connectivity $k_{out}$ are chosen from a power law distribution $P(k_{out}) \sim k_{out}^{-2}$, truncated such that $k_{out} \in [2,100]$.}
\end{figure}

In addition to the various avalanche distributions, we can also observe how the connectivity of the network evolves over time for different inhibitory fractions. Figure \ref{fig:init} shows an example of the initial distribution of outgoing connectivity for our networks. The horizontal axis represents the degree of outgoing connections $k_{out}$ for individual neurons, while the vertical axis represents the number of neurons measured with a particular value of $k_{out}$. The (blue) circles represent distribution of connectivity that was measured from our network upon its creation. The (orange) line is the best fit of a power law to the measured distribution, which follows a power law with exponent very close to $-2.0$. 

The data very clearly displays the behavior described in Sec. \ref{section:model}, which requires that the initial degrees of outgoing connectivity be drawn from a truncated power law distribution with exponent $-2.0$, constrained such that $k_{out}\in[2,100]$. While the following plots refer to specific inhibitory fractions, Fig. \ref{fig:init} serves as an example for all inhibitory fractions, as all networks are initialized in the same manner.   

As avalanches occur in the network, the Hebbian rules defined in Sec. \ref{section:model} will cause the the connections between neurons to change according to their use. Frequently used connection will be strengthened, and infrequently employed connections will be weakened or even ``pruned" (i.e., removed) if the connections become too small. We are interested in observing how the distribution of outgoing connectivity evolves as function of time and inhibitory fraction under these effects. 

\begin{figure*}
  \centering
  \begin{subfigure}{0.49\textwidth}
    \includegraphics[width=\textwidth]{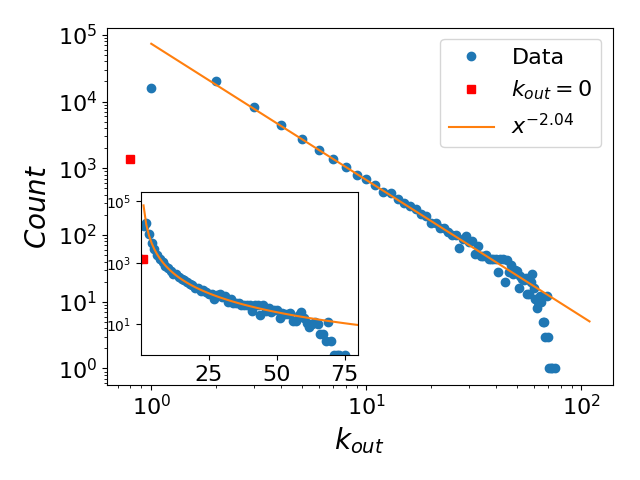}
    \caption{}
  \end{subfigure}
  \begin{subfigure}{0.49\textwidth}
    \includegraphics[width=\textwidth]{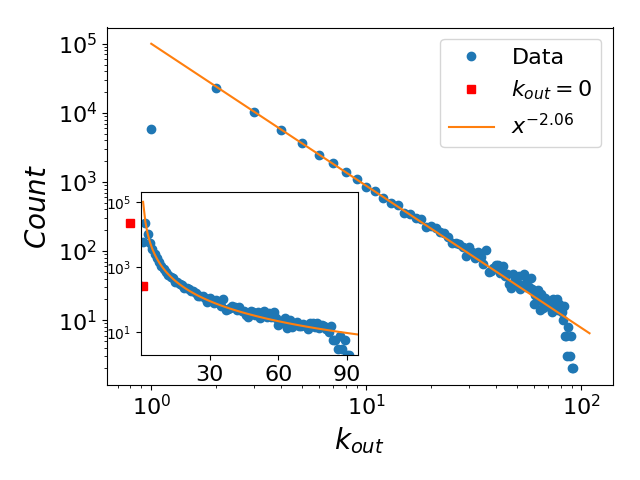}
    \caption{}
  \end{subfigure}
  \caption{\label{fig:high_kout} The $k_{out}$ degree distribution for networks with an inhibitory fraction of $p_{inh}=0.30$ after 45,000 avalanches. The horizontal axis ($k_{out}$) represents the number of outgoing connections individual neurons have. The vertical axis represents the number of neurons that have a particular value of $k_{out}$. The (blue) circles are the data points we measured and averaged from 100 networks with (a) $p_{inh}=0.30$, and (b) $p_{inh}=0.04$ after $45,000$ avalanches, and represent the number of neurons with each degree of connectivity except zero. The (red) squares represents the data points for the number of neurons with zero outgoing connections. These data points are plotted with some offset along the horizontal axis in order to display it on a double-logarithmic plot. The (orange) line is the best fit to the power law regime of the distribution, with exponents $-2.04$ (a) and $-2.06$ (b). The inset represents a log-linear graph of the same data, replotted to highlight the maximum value of $k_{out}$ for this network, which is $75$ (a) and $91$ (b), respectively.}
\end{figure*}

Figure \ref{fig:high_kout} shows the distribution of outgoing degrees of connectivity for two networks with (a) $p_{inh}=0.30$, and (b) $p_{inh}=0.04$ after $45,000$ avalanches. The horizontal axis again represents the degrees of outgoing connectivity $k_{out}$ of individual neurons, while the vertical axis displays the number of observed neurons with a particular $k_{out}$. The (blue) circles represent the number of neurons measured with all degrees of connectivity except zero, and the (red) square represents the measured number of neurons with zero outgoing connections. The (red) square is placed with some positive offset on the horizontal axis in order to display this data on a double-logarithmic plot. The (orange) line represents the best fit of a power law to the data, which follows an exponent very close to $-2.0$ in both situations. The inset is a log-linear graph of the same data, replotted to highlight the maximum degree of connectivity in the network, $k_{out} = 75$ for  $p_{inh}=0.04$ (a), and $k_{out}=91$ for $p_{inh}=0.04$ (b). While the power law behavior of the majority of the distribution is unchanged, there are striking differences in the head and tail of the distribution after the Hebbian rules have been applied for $45,000$ avalanches. Many of the high degrees of connectivity have been pruned out of the distribution, and many more neurons retain either only one or zero outgoing connections; additionally, no neurons have more than $75$ and $91$ outgoing connections, respectively, for these two inhibitory fractions.

Even with the general decrease in connectivity, the lower inhibitory fraction network is inherently more strongly connected than the higher inhibitory fraction network after a long simulation time allowing for $45,000$ avalanches. The neural network with smaller $p_{inh}$ displays a higher maximum value of $k_{out}$, and the number of neurons with only one or zero outgoing connections is reduced by almost an order of magnitude as compared with the network with larger $p_{inh}$. 

This difference in connectivity evolution results from the distinct inhibitory fractions of the networks as follows: A stronger inhibitory network will result in weaker signals being transmitted through it. Weaker signals in turn cause weaker connections between neurons, and hence an increase in the number of neuron connections pruned due to the Hebbian rules that govern the system. In comparison, the network with a lower inhibitory fraction cannot suppress the signals in the network as strongly as the network with a higher $p_{inh}$. The network with a lower inhibitory fraction will sustain stronger connections and will consequently not prune away neuron links as drastically. In general we observe that the pruning mechanism preserves the power law structure of the initial distribution, and the effects of the pruning are only visible in the very head and tails of the distribution.

\subsection{Control of avalanche distributions}
Figure \ref{fig:strength} shows how the avalanche strength distribution changes as the inhibitory fraction of the network is varied. The plot shows two regimes of activity: At higher inhibitory fractions the distribution follows a power law behavior terminating in an exponential cut-off; at very low inhibitory fractions the algebraic decay extends further and the exponential cut-off is shifted to very high avalanche strengths. These extended power laws dominate the dynamics of the networks they occur in. In the following subsection we draw inspiration from work previously done on the robustness of scale-free networks \cite{albert2000, Complex_Network} and propose two different control strategies to remove these extended power law tails in networks with low inhibitory fractions through the disabling of either (1) randomly picked or (2) specifically selected highly connected excitatory neurons. 

\subsubsection{Control through disabling random excitatory neurons}
The first strategy we implemented is disabling randomly chosen excitatory neurons. Disabling a neuron means that the internal potential of the neuron is permanently set to zero from the time step it is disabled. We choose excitatory neurons only because the goal is to prevent the large avalanches occurring in the extended tail of this network's normal avalanche strength distribution. An active inhibitory neuron is more effective at stopping these avalanches than a disabled neuron, so we only pick excitatory neurons. The neurons are selected with equal probability from all excitatory neurons. We tested several different fractions of excitatory neurons to disable randomly, and only observed the extended tail present in the avalanche strength distribution to disappear for disabling fractions greater or equal to $0.30$. 

\begin{figure*}[htp]
  \begin{subfigure}[b]{0.49\textwidth}
    \includegraphics[width=\textwidth]{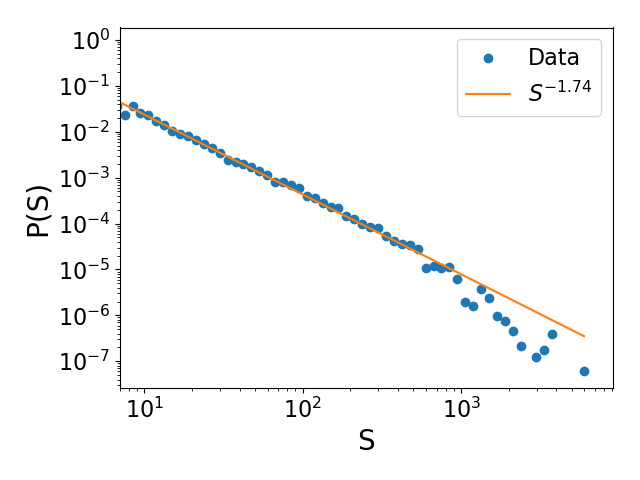}
    \caption{\label{fig:disable_random}}
  \end{subfigure}
  \begin{subfigure}[b]{0.49\textwidth}
    \includegraphics[width=\textwidth]{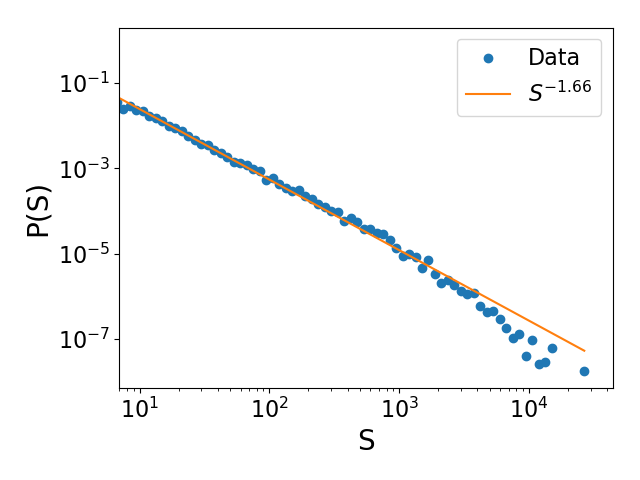}
    \caption{\label{fig:disable_targeted}}
  \end{subfigure}
  \caption{\label{fig:disable} The avalanche strength distribution of a network with an inhibitory fraction of $p_{inh}=0.04$ after (a) randomly selected $30\%$ of the excitatory neurons, and (b) the top $1\%$ of highly connected excitatory neurons have been disabled. In either case, the network was allowed to evolve naturally for the duration of $60,000$ avalanches before the excitatory neurons were disabled. The (blue) circles represent the probability of an avalanche having a given avalanche strength $S$, and the (orange) line is the best fit of a power law to the data. The extended tail present in a normal $p_{inh}=0.04$ network has disappeared after disabling (a) a large fraction, (b) a minor fraction of excitatory neurons. For case (a), disabling $30\%$ of inhibitory of neurons is very destructive to the dynamics of the network, and a marked change in the power law behavior of the avalanche strength distribution is observed: This network follows a power law with exponent $\sim -1.74$, lower than the normal exponent of $-1.5$. This is the result of signals dying out more quickly in this heavily diluted network. Disabling random fractions less than $0.30$ of the excitatory neurons retained the power law tails typically seen in the avalanche strength distributions of these low inhibitory fraction networks; indeed, $0.30$ is the lowest fraction of random excitatory neurons that must be disabled to curtail these extended power law tails.
    For (b), the extended tail present in a normal $p_{inh}=0.04$ network has disappeared after disabling only a minor fraction of the excitatory neurons. Due to the power law structure of the network connectivity, these top $1\%$ of the excitatory neurons are much more important to the network dynamics than the vast majority of all other neurons. Disabling these  neurons is still destructive to the network dynamics, though less so than the random disabling case shown in (a), and a change in the power law behavior of the network is observed. This network follows a power law with exponent $\sim -1.66$, again lower than the normal exponent of $-1.5$. This is the result of signals being unable to propagate through the most heavily connected neurons, which play a large role in the network's transmission capability. $0.01$ is the required lowest fraction of highly connected excitatory neurons that needs to be disabled in order to effectively remove these extended power law tails.
    Data averaged over $100$ independent network realizations.}
\end{figure*}

Figure \ref{fig:disable}(a) shows the avalanche strength distribution of a network with inhibitory fraction of $p_{inh}=0.04$ after $30\%$ of the excitatory neurons were randomly selected and disabled, i.e. $n_{i}(t)$ was held at zero for all disabled neurons in the subsequent evolution. The network was allowed to evolve unperturbed for $60,000$ avalanches before the excitatory neurons were disabled. The data shown was averaged over 100 realizations of the network. The extended tail seen in avalanche strength distributions with this inhibitory neuron fraction has disappeared, due to the fragmentation of the network caused by disabling so many neurons. The network is no longer able to sustain the large network-wide avalanches necessary for the extended power law tail in the avalanche strength distribution. Disabling fractions of random excitatory neurons less than $30\%$ does not remove the extended tail. The system is thus quite robust against random disablings, which is not surprising given its scale-free structure. Scale-free networks are known to be quite stable against the random removal of nodes, which is analogous to our disabling of neurons. In order to prevent a signal from being propagated across a generic scale-free network of the same size as our network, more than 
$90\%$ of the nodes must be randomly removed \cite{albert2000,Complex_Network}. We only need to disable a much lower fraction of random neurons to see the extended power law disappear because we are not attempting to disrupt the entire network activity, but only curtail the power law regime of these extended avalanches.  

However, even with the network's robustness, disabling this many neurons from the network is destructive to the normal dynamics of the network, and we observe a change in exponent of the power law behavior of the distribution from $-1.5$ to $-1.74$. This lower exponent results in a much lower probability of strong avalanches, and we see the cut-off appear several decades below the threshold in networks with higher inhibitory fractions. 

Disabling random neurons consequently is an ultimately successful strategy for removing the long power law tail in this avalanche distribution; however, it requires a significant portion of the network's neurons be disabled, which is certainly not ideal, and very likely quite detrimental in any biological neural network.

\subsubsection{Control through disabling highly connected neurons}
The second strategy we implemented is the disabling of the most highly connected excitatory neurons. The neurons were chosen based on their degree of outgoing connectivity $k_{out}$. We tested several different fractions of the most connected excitatory neurons, and observed that only the top $1\%$ of highly connected neurons need to be disabled in order to stop these incredibly large avalanches.

Figure \ref{fig:disable}(b) depicts the avalanche strength of a network with inhibitory fraction of $p_{inh}=0.04$ with the top $1\%$ of the most highly connected neurons disabled. The network was again allowed to evolve for $60,000$ avalanches before the highly connected neurons were disabled. The (blue) circles are the averaged data points of the avalanche strength distributions from $100$ different network realizations. The (orange) line is the best fit of a power law to the data. This model is very sensitive to disabling highly connected neurons, which is also to be expected given its scale-free structure. In addition to being very robust against random disablings, scale-free networks are highly susceptible to ``targeted" disablings, where the most highly connected nodes are removed \cite{albert2000,Complex_Network}. A generic scale-free network of the same size and exponent of the connectivity distribution needs approximately the top $3\%$ of highly connected nodes disabled to completely fragment the network and destroy any long-range connectivity \cite{albert2000,Complex_Network,Callaway2000}. In our system, we need only disable the top $1\%$ of the highly connected neurons because we do not aim to completely destroy the network dynamics, yet merely wish prevent the occurrence of exceedingly strong avalanches.

Permanently disabling these highly connected neurons does still considerably affect the network dynamics. The avalanche strength distribution follows a power law with exponent $-1.66$ instead of the typical $-1.5$. This is due to the signals being unable to propagate as strongly through the network after the highly connected neurons have been disabled. These weaker avalanches are more likely to die out earlier than their counterparts in an unsuppressed network, resulting in a lower exponent and an earlier cut-off in the avalanche strength distribution.


Disabling only the highly connected neurons is hence demonstrably a very successful control strategy for removing these extended power law tails that dominate the network. However, this approach does require significant knowledge about the structure of the network and is still destructive to the network dynamics because, by definition, these highly connected neurons are very important to signal propagation through the system.

\section{Discussion}
Operating our model with the parameters described in Sec. \ref{Neuron_dynamics} and an inhibitory fraction of $p_{inh}=0.30$, we observe that the avalanche strength distribution of our model follows a power law of $P_S(S) \sim S^{-1.55}$, and that the avalanche duration distribution of our model obeys a power law of $P_D(D) \sim D^{-2.1}$, both of which agree well with experimental results \cite{Beggs_Plenz_2003,Yu2014} and reproduce the results shown by Lombardi, Herrmann, De Arcangelis et al. As we lower the inhibitory fraction of our network towards an inhibitory fraction of $0.04$, we intriguingly find behavior suggestive of criticality as the exponential cut-off present previously in the avalanche strength and duration distributions disappears, and these distributions continue to follow power laws for several more decades. At this value of inhibitory fraction, the network becomes dominated by long-lasting avalanches that persist for billions of time steps. The particular value of inhibitory fraction at which we see this extension of the distributions is far below the fraction found in human cortices, which is closer to $0.2-0.3$ \cite{Sahara2012,deArcangelis_Power_Spectrum}. 

Additionally, the power spectral density of our network at low inhibitory fractions ($p_{inh}=0.04$) behaves similarly to power spectral densities of epileptic humans by following a power law with exponent $-2.0$. As the inhibitory fraction of the network is increased to a more biologically relevant value of $0.3$ the exponent of this power law increased to $-1.6$, which more closely matches the observed exponent of healthy human brains \cite{Yan2016}. The transition of the exponent between ``epileptic'' and ``healthy'' regimes   reproduces results observed by Lombardi, Herrmann, De Arcangelis et al.

Low inhibitory fractions allow the network to access much higher avalanche durations and corresponding avalanche strengths, because even though the underlying power law distribution of the these quantities does not change, the exponential cut-off disappears allowing the power law distributions to extends into regimes of greatly increased duration and strength. The incredibly large ``black swan events" that the network can access have correspondingly low probabilities due to the power law distribution, but because they are so large, they dominate the network for billions of time steps once they occur. 

The exponential cut-offs protect the network from these events, and human cortices may naturally operate at higher inhibitory fractions in order to avoid a truly critical point, yet still benefit from wide distributions at lower intensity avalanche events.

This corroborates the idea proposed by Milton et al. \cite{Milton2012}, that critical behavior in the brain, though long sought after, might be destructive as the long-range correlations introduced by approaching a critical point could destroy and dominate the short-range interactions necessary for the proper operation of the brain.

We also observe how the outgoing connectivity distribution changes as the network evolves under the Hebbian learning rules described in Sec. \ref{section:model} after $45,000$ avalanches have run through the system. 

Networks with a high inhibitory fraction ($p_{inh}=0.30$) prune away many connections, as the system is unable to propagate avalanches strong enough to sustain all of the links. This results in the tail of the connectivity distribution being truncated with ultimately no neurons maintaining more than $75$ outgoing connections, while the head of the distribution becomes inflated, as many neurons end up having only one or zero outgoing connections. Networks with a lower inhibitory fraction ($p_{inh}=0.04$), prune away fewer connections than networks with a higher inhibitory fraction, as they are able to sustain stronger avalanches in the network. This results in a extended connectivity distribution tail, with some neurons having as many as $91$ outgoing connections after $45,000$ avalanches. Additionally these networks display an order of magnitude fewer neurons with zero or merely one connection than the networks endowed with a higher inhibitory fraction.

The combination of the inhibitory neurons and the Hebbian rules of the system cause networks with high inhibitory fractions to evolve into more sparsely connected networks than networks with a low inhibitory fraction. These differences in connectivity reinforce the networks' ability to sustain or disrupt very large avalanches. Networks with high inhibitory fractions will display weaker avalanches, causing them to be less connected, which in turn further weakens them in their capability to sustain large-scale avalanches. Networks with lower inhibitory fractions will on occasion go through massive avalanches allowing them to remain more connected, which hence will assist these systems in permitting further strong avalanches.

Finally we investigate two different strategies to remove these exceedingly large avalanches from networks with low inhibitory fraction through either the disabling of randomly selected or carefully chosen highly connected excitatory neurons, respectively. In order to curtail these large events through random disablings, $30\%$ of the networks excitatory neurons must be disabled. This strategy is therefore ultimately effective, but would be quite destructive to any biological neural networks. In contrast, switching off highly connected neurons proves to be a much more effective strategy, as only the top $1\%$ of these prominently connected excitatory neurons need to be disabled in order to prevent such large-scale avalanche events. Both of these strategies provide a means to circumvent the inherent occurrence of  incredibly large ``epileptic" avalanches in systems with very low inhibitory neuron fraction.

\section*{Acknowledgements}
We would like to thank Michel Pleimling and Priyanka for helpful discussions, Alexandra Bosh for her contributions to this work in the course of her undergraduate research project, and Lucilla De Arcangelis, Hans J. Herrmann, and Fabricio Lombardi for valuable comments.
Research was sponsored by the Army Research Office and was accomplished under Grant Number W911NF-17-1-0156. The views and conclusions contained in this document are those of the authors and should not be interpreted as representing the official policies, either expressed or implied, of the Army Research Office or the U.S. Government. The U.S. Government is authorized to reproduce and distribute reprints for Government purposes notwithstanding any copyright notation herein.

\appendix*
\section{Extended model to reproduce waiting time distribution}
\label{Appendix:Arc_Model}
The model described in Sec. \ref{section:model} accurately recreates the avalanche strength distribution, the avalanche duration distribution, and the power spectrum distribution observed in rat and human cortices \cite{Beggs_Plenz_2003, Myers2017}. In order to also generate experimentally detected waiting time distributions, the model must be extended. This extended model is detailed by Lombardi, Herrmann, de Arcangelis et al. \cite{deArcangelis_BalanceExcInh, deArcangelis_DragonKing, deArcangelis_TemporalOrg,deArcangelis_Power_Spectrum} and is briefly summarized here.

\begin{figure}[htp]
  \centering
  \includegraphics[width=.5\textwidth]{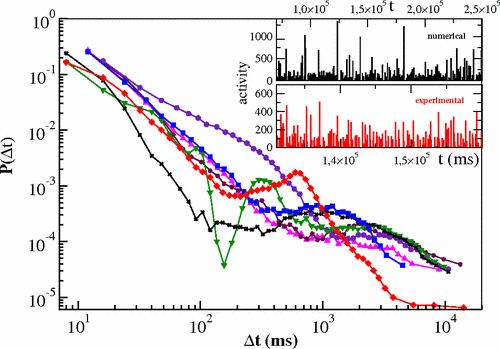}
  \caption{\label{fig:arc_wait_time} (Figure reproduced with permission from Ref. \cite{deArcangelis_BalanceExcInh}.) This plot shows the avalanche waiting time distributions for seven different slices of rat cortex. The inset shows two examples of temporal neural sequences. The majority of the waiting time distributions display bimodal behavior, with an initial power law at low waiting times, followed by an Gaussian bump at higher waiting times. }
\end{figure}

The waiting time between avalanches is the time between the end of the last avalanche and the beginning of the next. Figure \ref{fig:arc_wait_time} shows experimentally determined waiting time of seven different slices of rat cortex. This figure was taken with permission from Lombardi, Herrmann, de Arcangelis et al.'s paper {\it The balance between excitation and inhibition controls the temporal organization of neuronal avalanches} \cite{deArcangelis_BalanceExcInh}. 

The experimentally determined waiting time distributions \cite{deArcangelis_BalanceExcInh} shown in Fig. \ref{fig:arc_wait_time} display bimodal behavior with an initial power law regime followed by a Gaussian ``bump".

This bimodal behavior requires that avalanches which occur within short waiting times should be highly correlated to previous avalanches in order to reproduce the power law behavior at low waiting times. In comparison, avalanches with long waiting times need to be uncorrelated to reproduce the exponential behavior seen at longer waiting times.

These distinct features can be reproduced by introducing two network-wide macro-states: ``up" and ``down". The up state is defined as a period of high network activity, during which many neurons are close to firing potential. Anytime an avalanche occurs, the system transitions to or remains in the up state.

During an up state, the noise driving the system is drawn from the distribution $(0, S_{max}/S(\tau)]$, where $S_{max}$ is the Avalanche strength threshold, and $S(\tau)$ is the strength of the last avalanche. Additionally, when an avalanche ends in the up state, each neuron in the network is reset to be close to the neuron firing threshold.
  \begin{equation}
    \label{eq:neuron_resetting}
    n_{i} \rightarrow n_{max}(1-S(\tau)/S_{max})\; .
  \end{equation}

  The resetting in Eq. (\ref{eq:neuron_resetting}) ensures with high probability that avalanches in the up state will have short waiting times, and the correlations introduced from the up state noise distribution produces the power law behavior seen in Fig. \ref{fig:arc_wait_time}.

  If the strength of the last avalanche exceeds the avalanche strength threshold $S_{max}$, the system transitions to the down state. The down state is characterized as a period of no activity in the network in which the system is slowly brought back to firing. When the network transitions from the up state to the down state, each neuron is drastically polarized in opposition of the previous behavior,

  \begin{equation}
    n_{i} \rightarrow n_{i} - h\Delta n_{i}\; ,
  \end{equation}
  where $\Delta n_{i}$ is the sum of depolarizations during the last avalanche in the up state, and $h$ is a system parameter introduced to control the strength at which the neuron is anti-polarized. During the down state, the network is driven by small ($\sim 0.01\cdot n_{max}$) constant noise.

  The hyperpolarization of neurons coupled with the small constant noise ensures that the waiting times during the down state will be very long, and will produce an approximately Gaussian distribution due to the central limit theorem.

  \bibliography{netBib}

\begin{thebibliography}{22}%
\makeatletter
\providecommand \@ifxundefined [1]{%
 \@ifx{#1\undefined}
}%
\providecommand \@ifnum [1]{%
 \ifnum #1\expandafter \@firstoftwo
 \else \expandafter \@secondoftwo
 \fi
}%
\providecommand \@ifx [1]{%
 \ifx #1\expandafter \@firstoftwo
 \else \expandafter \@secondoftwo
 \fi
}%
\providecommand \natexlab [1]{#1}%
\providecommand \enquote  [1]{``#1''}%
\providecommand \bibnamefont  [1]{#1}%
\providecommand \bibfnamefont [1]{#1}%
\providecommand \citenamefont [1]{#1}%
\providecommand \href@noop [0]{\@secondoftwo}%
\providecommand \href [0]{\begingroup \@sanitize@url \@href}%
\providecommand \@href[1]{\@@startlink{#1}\@@href}%
\providecommand \@@href[1]{\endgroup#1\@@endlink}%
\providecommand \@sanitize@url [0]{\catcode `\\12\catcode `\$12\catcode
  `\&12\catcode `\#12\catcode `\^12\catcode `\_12\catcode `\%12\relax}%
\providecommand \@@startlink[1]{}%
\providecommand \@@endlink[0]{}%
\providecommand \url  [0]{\begingroup\@sanitize@url \@url }%
\providecommand \@url [1]{\endgroup\@href {#1}{\urlprefix }}%
\providecommand \urlprefix  [0]{URL }%
\providecommand \Eprint [0]{\href }%
\providecommand \doibase [0]{http://dx.doi.org/}%
\providecommand \selectlanguage [0]{\@gobble}%
\providecommand \bibinfo  [0]{\@secondoftwo}%
\providecommand \bibfield  [0]{\@secondoftwo}%
\providecommand \translation [1]{[#1]}%
\providecommand \BibitemOpen [0]{}%
\providecommand \bibitemStop [0]{}%
\providecommand \bibitemNoStop [0]{.\EOS\space}%
\providecommand \EOS [0]{\spacefactor3000\relax}%
\providecommand \BibitemShut  [1]{\csname bibitem#1\endcsname}%
\let\auto@bib@innerbib\@empty
\bibitem [{\citenamefont {Fain}(2014)}]{Neuron_Biology}%
  \BibitemOpen
  \bibfield  {author} {\bibinfo {author} {\bibfnamefont {G.~L.}\ \bibnamefont
  {Fain}},\ }\href@noop {} {\emph {\bibinfo {title} {Molecular and Cellular
  Physiology of Neurons}}},\ \bibinfo {edition} {2nd}\ ed.\ (\bibinfo
  {publisher} {Harvard University Press Cambridge},\ \bibinfo {year} {2014})\
  p.\ \bibinfo {pages} {178}\BibitemShut {NoStop}%
\bibitem [{\citenamefont {Schmidt}\ and\ \citenamefont
  {Thews}(1989)}]{Human_Physiology}%
  \BibitemOpen
  \bibfield  {author} {\bibinfo {author} {\bibfnamefont {R.~F.}\ \bibnamefont
  {Schmidt}}\ and\ \bibinfo {author} {\bibfnamefont {G.}~\bibnamefont
  {Thews}},\ }\href@noop {} {\emph {\bibinfo {title} {Human Physiology}}},\
  \bibinfo {edition} {2nd}\ ed.\ (\bibinfo  {publisher} {Springer Berlin
  Heidelberg},\ \bibinfo {year} {1989})\ p.~\bibinfo {pages} {26}\BibitemShut
  {NoStop}%
\bibitem [{\citenamefont {Gaiarsa}\ \emph {et~al.}(2002)\citenamefont
  {Gaiarsa}, \citenamefont {Caillard},\ and\ \citenamefont
  {Ben-Ari}}]{Gaiarsa2002}%
  \BibitemOpen
  \bibfield  {author} {\bibinfo {author} {\bibfnamefont {J.-L.}\ \bibnamefont
  {Gaiarsa}}, \bibinfo {author} {\bibfnamefont {O.}~\bibnamefont {Caillard}}, \
  and\ \bibinfo {author} {\bibfnamefont {Y.}~\bibnamefont {Ben-Ari}},\ }\href
  {\doibase 10.1016/S0166-2236(02)02269-5} {\bibfield  {journal} {\bibinfo
  {journal} {Trends in Neurosciences}\ }\textbf {\bibinfo {volume} {25}},\
  \bibinfo {pages} {564 } (\bibinfo {year} {2002})}\BibitemShut {NoStop}%
\bibitem [{\citenamefont {Gerrow}\ and\ \citenamefont
  {Triller}(2010)}]{Gerrow2010}%
  \BibitemOpen
  \bibfield  {author} {\bibinfo {author} {\bibfnamefont {K.}~\bibnamefont
  {Gerrow}}\ and\ \bibinfo {author} {\bibfnamefont {A.}~\bibnamefont
  {Triller}},\ }\href {\doibase 10.1016/j.conb.2010.06.010} {\bibfield
  {journal} {\bibinfo  {journal} {Current Opinion in Neurobiology}\ }\textbf
  {\bibinfo {volume} {20}},\ \bibinfo {pages} {631 } (\bibinfo {year}
  {2010})},\ \bibinfo {note} {neuronal and glial cell biology – New
  technologies}\BibitemShut {NoStop}%
\bibitem [{\citenamefont {Malenka}\ and\ \citenamefont
  {Bear}(2004)}]{Malenka2004}%
  \BibitemOpen
  \bibfield  {author} {\bibinfo {author} {\bibfnamefont {R.~C.}\ \bibnamefont
  {Malenka}}\ and\ \bibinfo {author} {\bibfnamefont {M.~F.}\ \bibnamefont
  {Bear}},\ }\href {\doibase 10.1016/j.neuron.2004.09.012} {\bibfield
  {journal} {\bibinfo  {journal} {Neuron}\ }\textbf {\bibinfo {volume} {44}},\
  \bibinfo {pages} {5 } (\bibinfo {year} {2004})}\BibitemShut {NoStop}%
\bibitem [{\citenamefont {Beggs}\ and\ \citenamefont
  {Plenz}(2003)}]{Beggs_Plenz_2003}%
  \BibitemOpen
  \bibfield  {author} {\bibinfo {author} {\bibfnamefont {J.~M.}\ \bibnamefont
  {Beggs}}\ and\ \bibinfo {author} {\bibfnamefont {D.}~\bibnamefont {Plenz}},\
  }\href {\doibase 10.1523/JNEUROSCI.23-35-11167.2003} {\bibfield  {journal}
  {\bibinfo  {journal} {Journal of Neuroscience}\ }\textbf {\bibinfo {volume}
  {23}},\ \bibinfo {pages} {11167} (\bibinfo {year} {2003})},\ \Eprint
  {http://arxiv.org/abs/http://www.jneurosci.org/content/23/35/11167.full.pdf}
  {http://www.jneurosci.org/content/23/35/11167.full.pdf} \BibitemShut
  {NoStop}%
\bibitem [{\citenamefont {Yan}\ \emph {et~al.}(2016)\citenamefont {Yan},
  \citenamefont {Wang}, \citenamefont {Ouyang}, \citenamefont {Yu},
  \citenamefont {Li}, \citenamefont {Sik},\ and\ \citenamefont {Li}}]{Yan2016}%
  \BibitemOpen
  \bibfield  {author} {\bibinfo {author} {\bibfnamefont {J.}~\bibnamefont
  {Yan}}, \bibinfo {author} {\bibfnamefont {Y.}~\bibnamefont {Wang}}, \bibinfo
  {author} {\bibfnamefont {G.}~\bibnamefont {Ouyang}}, \bibinfo {author}
  {\bibfnamefont {T.}~\bibnamefont {Yu}}, \bibinfo {author} {\bibfnamefont
  {Y.}~\bibnamefont {Li}}, \bibinfo {author} {\bibfnamefont {A.}~\bibnamefont
  {Sik}}, \ and\ \bibinfo {author} {\bibfnamefont {X.}~\bibnamefont {Li}},\
  }\href {\doibase 10.1007/s11071-015-2455-9} {\bibfield  {journal} {\bibinfo
  {journal} {Nonlinear Dynamics}\ }\textbf {\bibinfo {volume} {83}},\ \bibinfo
  {pages} {1909} (\bibinfo {year} {2016})}\BibitemShut {NoStop}%
\bibitem [{\citenamefont {Yu}\ \emph {et~al.}(2014)\citenamefont {Yu},
  \citenamefont {Klaus}, \citenamefont {Yang},\ and\ \citenamefont
  {Plenz}}]{Yu2014}%
  \BibitemOpen
  \bibfield  {author} {\bibinfo {author} {\bibfnamefont {S.}~\bibnamefont
  {Yu}}, \bibinfo {author} {\bibfnamefont {A.}~\bibnamefont {Klaus}}, \bibinfo
  {author} {\bibfnamefont {H.}~\bibnamefont {Yang}}, \ and\ \bibinfo {author}
  {\bibfnamefont {D.}~\bibnamefont {Plenz}},\ }\href
  {10.1371/journal.pone.0099761} {\bibfield  {journal} {\bibinfo  {journal}
  {PLOS ONE}\ }\textbf {\bibinfo {volume} {9}},\ \bibinfo {pages} {1} (\bibinfo
  {year} {2014})}\BibitemShut {NoStop}%
\bibitem [{\citenamefont {Lombardi}\ \emph {et~al.}(2012)\citenamefont
  {Lombardi}, \citenamefont {Herrmann}, \citenamefont {Perrone-Capano},
  \citenamefont {Plenz},\ and\ \citenamefont
  {De~Arcangelis}}]{deArcangelis_BalanceExcInh}%
  \BibitemOpen
  \bibfield  {author} {\bibinfo {author} {\bibfnamefont {F.}~\bibnamefont
  {Lombardi}}, \bibinfo {author} {\bibfnamefont {H.~J.}\ \bibnamefont
  {Herrmann}}, \bibinfo {author} {\bibfnamefont {C.}~\bibnamefont
  {Perrone-Capano}}, \bibinfo {author} {\bibfnamefont {D.}~\bibnamefont
  {Plenz}}, \ and\ \bibinfo {author} {\bibfnamefont {L.}~\bibnamefont
  {De~Arcangelis}},\ }\href {\doibase 10.1103/PhysRevLett.108.228703}
  {\bibfield  {journal} {\bibinfo  {journal} {Physical Review Letters}\
  }\textbf {\bibinfo {volume} {108}},\ \bibinfo {pages} {228703} (\bibinfo
  {year} {2012})}\BibitemShut {NoStop}%
\bibitem [{\citenamefont {De~Arcangelis}(2012)}]{deArcangelis_DragonKing}%
  \BibitemOpen
  \bibfield  {author} {\bibinfo {author} {\bibfnamefont {L.}~\bibnamefont
  {De~Arcangelis}},\ }\href {\doibase 10.1140/epjst/e2012-01574-6} {\bibfield
  {journal} {\bibinfo  {journal} {The European Physical Journal Special
  Topics}\ }\textbf {\bibinfo {volume} {205}},\ \bibinfo {pages} {243}
  (\bibinfo {year} {2012})}\BibitemShut {NoStop}%
\bibitem [{\citenamefont {Lombardi}\ and\ \citenamefont
  {De~Arcangelis}(2014)}]{deArcangelis_TemporalOrg}%
  \BibitemOpen
  \bibfield  {author} {\bibinfo {author} {\bibfnamefont {F.}~\bibnamefont
  {Lombardi}}\ and\ \bibinfo {author} {\bibfnamefont {L.}~\bibnamefont
  {De~Arcangelis}},\ }\href {\doibase 10.1140/epjst/e2014-02253-4} {\bibfield
  {journal} {\bibinfo  {journal} {The European Physical Journal Special
  Topics}\ }\textbf {\bibinfo {volume} {223}},\ \bibinfo {pages} {2119}
  (\bibinfo {year} {2014})}\BibitemShut {NoStop}%
\bibitem [{\citenamefont {Lombardi}\ \emph {et~al.}(2017)\citenamefont
  {Lombardi}, \citenamefont {Herrmann},\ and\ \citenamefont
  {De~Arcangelis}}]{deArcangelis_Power_Spectrum}%
  \BibitemOpen
  \bibfield  {author} {\bibinfo {author} {\bibfnamefont {F.}~\bibnamefont
  {Lombardi}}, \bibinfo {author} {\bibfnamefont {H.~J.}\ \bibnamefont
  {Herrmann}}, \ and\ \bibinfo {author} {\bibfnamefont {L.}~\bibnamefont
  {De~Arcangelis}},\ }\href {\doibase 10.1063/1.4979043} {\bibfield  {journal}
  {\bibinfo  {journal} {Chaos: An Interdisciplinary Journal of Nonlinear
  Science}\ }\textbf {\bibinfo {volume} {27}},\ \bibinfo {pages} {047402}
  (\bibinfo {year} {2017})},\ \Eprint
  {http://arxiv.org/abs/https://doi.org/10.1063/1.4979043}
  {https://doi.org/10.1063/1.4979043} \BibitemShut {NoStop}%
\bibitem [{\citenamefont {Sahara}\ \emph {et~al.}(2012)\citenamefont {Sahara},
  \citenamefont {Yanagawa}, \citenamefont {O{\textquoteright}Leary},\ and\
  \citenamefont {Stevens}}]{Sahara2012}%
  \BibitemOpen
  \bibfield  {author} {\bibinfo {author} {\bibfnamefont {S.}~\bibnamefont
  {Sahara}}, \bibinfo {author} {\bibfnamefont {Y.}~\bibnamefont {Yanagawa}},
  \bibinfo {author} {\bibfnamefont {D.~D.~M.}\ \bibnamefont
  {O{\textquoteright}Leary}}, \ and\ \bibinfo {author} {\bibfnamefont {C.~F.}\
  \bibnamefont {Stevens}},\ }\href {\doibase 10.1523/JNEUROSCI.6412-11.2012}
  {\bibfield  {journal} {\bibinfo  {journal} {Journal of Neuroscience}\
  }\textbf {\bibinfo {volume} {32}},\ \bibinfo {pages} {4755} (\bibinfo {year}
  {2012})},\ \Eprint
  {http://arxiv.org/abs/http://www.jneurosci.org/content/32/14/4755.full.pdf}
  {http://www.jneurosci.org/content/32/14/4755.full.pdf} \BibitemShut {NoStop}%
\bibitem [{\citenamefont {Cooper}(2005)}]{Cooper_Hebb}%
  \BibitemOpen
  \bibfield  {author} {\bibinfo {author} {\bibfnamefont {S.~J.}\ \bibnamefont
  {Cooper}},\ }\href {\doibase 10.1016/j.neubiorev.2004.09.009} {\bibfield
  {journal} {\bibinfo  {journal} {Neuroscience \& Biobehavioral Reviews}\
  }\textbf {\bibinfo {volume} {28}},\ \bibinfo {pages} {851 } (\bibinfo {year}
  {2005})}\BibitemShut {NoStop}%
\bibitem [{\citenamefont {Priezzhev}\ \emph {et~al.}(1996)\citenamefont
  {Priezzhev}, \citenamefont {Ktitarev},\ and\ \citenamefont
  {Ivashkevich}}]{Sandpile}%
  \BibitemOpen
  \bibfield  {author} {\bibinfo {author} {\bibfnamefont {V.~B.}\ \bibnamefont
  {Priezzhev}}, \bibinfo {author} {\bibfnamefont {D.~V.}\ \bibnamefont
  {Ktitarev}}, \ and\ \bibinfo {author} {\bibfnamefont {E.~V.}\ \bibnamefont
  {Ivashkevich}},\ }\href {\doibase 10.1103/PhysRevLett.76.2093} {\bibfield
  {journal} {\bibinfo  {journal} {Physical Review Letters}\ }\textbf {\bibinfo
  {volume} {76}},\ \bibinfo {pages} {2093} (\bibinfo {year}
  {1996})}\BibitemShut {NoStop}%
\bibitem [{\citenamefont {{Egu{\'{\i}}luz}}\ \emph {et~al.}(2005)\citenamefont
  {{Egu{\'{\i}}luz}}, \citenamefont {{Chialvo}}, \citenamefont {{Cecchi}},
  \citenamefont {{Baliki}},\ and\ \citenamefont {{Apkarian}}}]{Eguiluz2005}%
  \BibitemOpen
  \bibfield  {author} {\bibinfo {author} {\bibfnamefont {V.~M.}\ \bibnamefont
  {{Egu{\'{\i}}luz}}}, \bibinfo {author} {\bibfnamefont {D.~R.}\ \bibnamefont
  {{Chialvo}}}, \bibinfo {author} {\bibfnamefont {G.~A.}\ \bibnamefont
  {{Cecchi}}}, \bibinfo {author} {\bibfnamefont {M.}~\bibnamefont {{Baliki}}},
  \ and\ \bibinfo {author} {\bibfnamefont {A.~V.}\ \bibnamefont {{Apkarian}}},\
  }\href {\doibase 10.1103/PhysRevLett.94.018102} {\bibfield  {journal}
  {\bibinfo  {journal} {Physical Review Letters}\ }\textbf {\bibinfo {volume}
  {94}},\ \bibinfo {eid} {018102} (\bibinfo {year} {2005})},\ \Eprint
  {http://arxiv.org/abs/cond-mat/0309092} {cond-mat/0309092} \BibitemShut
  {NoStop}%
\bibitem [{\citenamefont {Lee}\ \emph {et~al.}(2010)\citenamefont {Lee},
  \citenamefont {Oh}, \citenamefont {Kim}, \citenamefont {Noh}, \citenamefont
  {Choi},\ and\ \citenamefont {Mashour}}]{Lee2010}%
  \BibitemOpen
  \bibfield  {author} {\bibinfo {author} {\bibfnamefont {U.}~\bibnamefont
  {Lee}}, \bibinfo {author} {\bibfnamefont {G.}~\bibnamefont {Oh}}, \bibinfo
  {author} {\bibfnamefont {S.}~\bibnamefont {Kim}}, \bibinfo {author}
  {\bibfnamefont {G.}~\bibnamefont {Noh}}, \bibinfo {author} {\bibfnamefont
  {B.}~\bibnamefont {Choi}}, \ and\ \bibinfo {author} {\bibfnamefont {G.~A.}\
  \bibnamefont {Mashour}},\ }\href {\doibase 10.1097/ALN.0b013e3181f229b5}
  {\bibfield  {journal} {\bibinfo  {journal} {Anesthesiology}\ }\textbf
  {\bibinfo {volume} {113}},\ \bibinfo {pages} {1081} (\bibinfo {year}
  {2010})}\BibitemShut {NoStop}%
\bibitem [{\citenamefont {Myers}\ \emph {et~al.}(2017)\citenamefont {Myers},
  \citenamefont {Jolly}, \citenamefont {Li}, \citenamefont {de~Jongh~Curry},\
  and\ \citenamefont {Parfenova}}]{Myers2017}%
  \BibitemOpen
  \bibfield  {author} {\bibinfo {author} {\bibfnamefont {M.~H.}\ \bibnamefont
  {Myers}}, \bibinfo {author} {\bibfnamefont {E.}~\bibnamefont {Jolly}},
  \bibinfo {author} {\bibfnamefont {Y.}~\bibnamefont {Li}}, \bibinfo {author}
  {\bibfnamefont {A.}~\bibnamefont {de~Jongh~Curry}}, \ and\ \bibinfo {author}
  {\bibfnamefont {H.}~\bibnamefont {Parfenova}},\ }\href {\doibase
  10.1159/000464418} {\bibfield  {journal} {\bibinfo  {journal} {Annals of
  Neurosciences}\ }\textbf {\bibinfo {volume} {24}},\ \bibinfo {pages} {12}
  (\bibinfo {year} {2017})}\BibitemShut {NoStop}%
\bibitem [{\citenamefont {Albert}\ \emph {et~al.}(2000)\citenamefont {Albert},
  \citenamefont {Jeong},\ and\ \citenamefont {Bar\'abasi}}]{albert2000}%
  \BibitemOpen
  \bibfield  {author} {\bibinfo {author} {\bibfnamefont {R.}~\bibnamefont
  {Albert}}, \bibinfo {author} {\bibfnamefont {H.}~\bibnamefont {Jeong}}, \
  and\ \bibinfo {author} {\bibfnamefont {A.}~\bibnamefont {Bar\'abasi}},\
  }\href@noop {} {\bibfield  {journal} {\bibinfo  {journal} {Nature}\ }\textbf
  {\bibinfo {volume} {406}},\ \bibinfo {pages} {378} (\bibinfo {year}
  {2000})}\BibitemShut {NoStop}%
\bibitem [{\citenamefont {Albert}\ and\ \citenamefont
  {Barab\'asi}(2002)}]{Complex_Network}%
  \BibitemOpen
  \bibfield  {author} {\bibinfo {author} {\bibfnamefont {R.}~\bibnamefont
  {Albert}}\ and\ \bibinfo {author} {\bibfnamefont {A.-L.}\ \bibnamefont
  {Barab\'asi}},\ }\href {\doibase 10.1103/RevModPhys.74.47} {\bibfield
  {journal} {\bibinfo  {journal} {Rev. Mod. Phys.}\ }\textbf {\bibinfo {volume}
  {74}},\ \bibinfo {pages} {47} (\bibinfo {year} {2002})}\BibitemShut {NoStop}%
\bibitem [{\citenamefont {Callaway}\ \emph {et~al.}(2000)\citenamefont
  {Callaway}, \citenamefont {Newman}, \citenamefont {Strogatz},\ and\
  \citenamefont {Watts}}]{Callaway2000}%
  \BibitemOpen
  \bibfield  {author} {\bibinfo {author} {\bibfnamefont {D.~S.}\ \bibnamefont
  {Callaway}}, \bibinfo {author} {\bibfnamefont {M.~E.~J.}\ \bibnamefont
  {Newman}}, \bibinfo {author} {\bibfnamefont {S.~H.}\ \bibnamefont
  {Strogatz}}, \ and\ \bibinfo {author} {\bibfnamefont {D.~J.}\ \bibnamefont
  {Watts}},\ }\href {\doibase 10.1103/PhysRevLett.85.5468} {\bibfield
  {journal} {\bibinfo  {journal} {Phys. Rev. Lett.}\ }\textbf {\bibinfo
  {volume} {85}},\ \bibinfo {pages} {5468} (\bibinfo {year}
  {2000})}\BibitemShut {NoStop}%
\bibitem [{\citenamefont {Milton}(2012)}]{Milton2012}%
  \BibitemOpen
  \bibfield  {author} {\bibinfo {author} {\bibfnamefont {J.~G.}\ \bibnamefont
  {Milton}},\ }\href {\doibase 10.1111/j.1460-9568.2012.08102.x} {\bibfield
  {journal} {\bibinfo  {journal} {European Journal of Neuroscience}\ }\textbf
  {\bibinfo {volume} {36}},\ \bibinfo {pages} {2156} (\bibinfo {year}
  {2012})}\BibitemShut {NoStop}%
\end{thebibliography}%
\end{document}